\documentclass[aps,12pt,pra,showpacs,superscriptaddress,preprint]{revtex4}

\usepackage{amsmath}

\usepackage[dvipdfmx]{graphicx}

\usepackage{subfig}

\usepackage{array}

\usepackage{amsfonts}

\usepackage{epsf}

\usepackage{epsfig}

\usepackage{amssymb}

\usepackage{bbm}

\usepackage{delarray}

\usepackage{caption}

\usepackage{amstext}

\usepackage{graphics}

\usepackage{epstopdf}

\usepackage{color}

\catcode`ð=\active

\defð{\u{g}}

\catcode`Ð=\active

\defÐ{\u{G}}

\catcode`Ý=\active

\defÝ{\. I}

\catcode`ö=\active

\defö{\"{o}}

\catcode`Ö=\active

\defÖ{\"O}

\catcode`ü=\active

\defü{\"{u}}

\catcode`Ü=\active

\defÜ{\"{U}}

\catcode`Þ=\active

\defÞ{\c{S}}

\catcode`þ=\active

\defþ{\c{s}}

\catcode`ý=\active

\defý{{\i}}

\catcode`ç=\active

\defç{\d{c}}

\catcode`Ç=\active

\defÇ{\d{C}}

\begin{document}

\title{Exact Solutions of Effective-Mass Dirac-Pauli Equation with an Electromagnetic Field}
\author{\small Altuð Arda}
\altaffiliation[Present adress: ]{Department of Mathematical Science, City University London,
Northampton Square,\\ London EC1V 0HB, UK}\affiliation{Department of
Physics Education, Hacettepe University, 06800, Ankara,Turkey}
\author{\small Ramazan Sever}
\email[E-mails: ]{arda@hacettepe.edu.tr, sever@metu.edu.tr}\affiliation{Department of
Physics, Middle East Technical  University, 06531, Ankara,Turkey}

\begin{abstract}
The exact bound state solutions of the Dirac-Pauli equation are studied for an appropriate position-dependent mass function by using the Nikiforov-Uvarov method. For a central electric field having a shifted inverse linear term, all two kinds of solutions for bound states are obtained in closed forms.
\end{abstract}


\maketitle

\newpage

\section{I\lowercase{ntroduction}}

Describing of spin-$\frac{1}{2}$ particles and their moving in an electromagnetic field are expressed by the Dirac equation. The interaction of the spin-$\frac{1}{2}$ particle with the electromagnetic field is usually constructed by using the covariant derivative, $D_{\mu}$, in four-momentum operator. This coupling  to electromagnetic fields is named as minimal electromagnetic interaction rule by Gell-Mann. Another option which describes the interaction due to the anomalous magnetic moment of a charged particle is possible by using a non-minimal coupling. The extended version of the Dirac equation by non-minimal coupling term is the Dirac-Pauli equation [1]. This equation can also describe interactions of neutral spin-$\frac{1}{2}$ particles with electromagnetic fields as well and this situation is especially interesting. It is possible to investigate the exact solutions of this equation for various electromagnetic field configurations, such as a constant magnetic field, an electromagnetic plane wave [2]. The study of the Dirac Hamiltonian for spin-$\frac{1}{2}$ particles with anomalous magnetic moment in electromagnetic fields has especially become important [3, 4], because a duality appears between the anomalous magnetic moment and the electric charge of electron [5] which is known as Aharonov-Casher effect [6].

In the present paper, we study the exact solutions of the Dirac-Pauli equation for a neutral spin-$\frac{1}{2}$ particle with magnetic moment $\mu$ moving in an electromagnetic field within the position-dependent mass formalism. The bound state problem for the Dirac-Pauli equation has not attracted enough attention in literature [7] where the closed form for the energy levels and the wave functions are given. In Ref. [8], the quasi-exact solutions of the Dirac-Pauli equation for a neutral particle in an electric field have been searched in spherical, cylindrical, and Cartesian coordinates, respectively. In this point of view, investigating the exact solutions of the Dirac-Pauli equation and giving the results in closed forms could be interesting. It is well known that the problems dealing with exact solutions  of non-relativistic and/or relativistic wave equations are an important part of quantum mechanics. The above circumstances are our basic motivation to present this work.

Among the mass distributions used in the literature [9], the mass functions having singularities are especially prized in the view of the present work. The reader can find such kinds of mass distributions in Refs. [10-13]. The mass distribution obtained below has a depending also on the energy of the particle, and will give us the opportunity such as solving the equations analytically. The dependence of the mass on the energy is an approximation scheme used in semiconductor theory, heterostructures [14, 15], and quantum dots [16]. The energy dependence of the mass is proposed within the Hartree mean-field formalism to obtain the global state properties and the single-nucleon levels [17].

The organization of the work as follows. In Section II, by using the method of separation of variables, we obtain four partial coupled differential equations (DE's) for the Dirac-Pauli equation of a neutral spin-$\frac{1}{2}$ particle in a central electric field. From stationary Dirac-Pauli equation can be obtained second order DE's for each of two radial wave functions which we solve by using the Nikiforov-Uvarov (NU) method. In Section III, we give a brief outline for the NU method. In Section IV, we find the energy eigenvalues and the radial wave functions by using NU method for a spherically symmetric electric field including a shifted inverse linear term while the mass depends on spatially coordinates. We observe that the bound state solutions are possible for such an electromagnetic configuration within the position-dependent mass formalism, and the degeneracy of a energy level is finite.

\section{T\lowercase{he} D\lowercase{irac}-P\lowercase{auli} E\lowercase{quation} a\lowercase{nd} S\lowercase{pherical} S\lowercase{ymmetry}}

Now let us consider a neutral spin-$\frac{1}{2}$ particle with mass $M$ moving in an external electromagnetic field described by the field strength $F_{\mu\nu}$. The four-component spinor $\Psi(t,\vec{r})$ describing this particle satisfies the Dirac-Pauli equation ($\hbar=c=1$) [1]
\begin{eqnarray}
\left(i\gamma^{\mu}\partial_{\mu}-\frac{1}{2}\,\mu\sigma^{\mu\nu}F_{\mu\nu}-M\right)\Psi(t,\vec{r})=0\,,
\end{eqnarray}
where $\gamma^{\mu}=(\gamma^{0}, \vec{\gamma})$ are the Dirac matrices obeying $\{\gamma^{\mu}, \gamma^{\nu}\}=2g^{\mu\nu}$ with $g^{\mu\nu}=diag(1, -1, -1, -1)$ and $\sigma^{\mu\nu}=\frac{i}{2}[\gamma^{\mu}, \gamma^{\nu}]$. The second term in Eq. (1) is written in terms of external electric field $\vec{E}$ and magnetic field $\vec{B}$ as following
\begin{eqnarray}
\frac{1}{2}\,\sigma^{\mu\nu}F_{\mu\nu}=i\vec{\alpha}.\vec{E}-\vec{\Sigma}.\vec{B}\,,
\end{eqnarray}
where $\vec{\alpha}=\gamma^{0}\vec{\gamma}$, and $\Sigma^{k}=\frac{1}{2}\,\epsilon^{ijk}\sigma^{ij}$ where $\epsilon^{ijk}$ is totally antisymmetric tensors ($\epsilon^{123}=1$). The four-component spinor $\Psi(t,\vec{r})$ can be given as
\begin{eqnarray}
\Psi(t,\vec{r})=e^{-i\varepsilon t}\psi(\vec{r})\,,
\end{eqnarray}
if the electric field $\vec{E}$, and the magnetic one $\vec{B}$ are time-independent. By inserting Eq. (3) into Eq. (1), one obtains a stationary Dirac-Pauli equation
\begin{eqnarray}
\left(\vec{\alpha}.\vec{p}+i\mu\vec{\gamma}.\vec{E}-\mu\beta\vec{\Sigma}.\vec{B}+\beta M\right)\psi(\vec{r})=\varepsilon\psi(\vec{r})\,,
\end{eqnarray}
where $\vec{p}=-i\nabla$ and $\beta=\gamma^{0}$. We use the following representation for the Dirac matrices [8]
\begin{eqnarray}
\vec{\alpha}=\begin{pmatrix}
0 & \vec{\sigma}\\
\vec{\sigma} & 0
\end{pmatrix}\,,\,\,\,\beta=\begin{pmatrix}
1 & 0 \\
0 & -1
\end{pmatrix}\,,\nonumber
\end{eqnarray}
where $\vec{\alpha}$ are the Pauli matrices. We define $\psi=(\phi, \chi)^{t}$, where $t$ denotes transpose, and $\phi$ and $\chi$ are two-component spinors, respectively. Let us consider the case where only electric fields exist. A brief discussion for the case where both electric and magnetic fields simultaneously presence can be found in Ref. [7]. The stationary Dirac-Pauli equation (4) then takes the form
\begin{subequations}
\begin{align}
(\varepsilon+M)\chi=\vec{\sigma}.(\vec{p}-i\mu\vec{E})\phi\,,\\
(\varepsilon-M)\phi=\vec{\sigma}.(\vec{p}+i\mu\vec{E})\chi\,,
\end{align}
\end{subequations}
where we have four coupled partial DE's.

Now let us define the following
\begin{eqnarray}
\varphi^{\eta}(\vec{x})=\left\{\begin{array}{rcl} \varphi^{+}(\vec{x}) & \mbox{for} & \ell=j-\frac{1}{2}\,,\\
\varphi^{-}(\vec{x}) & \mbox{for} & \ell=j+\frac{1}{2}\,,\end{array}\right.
\end{eqnarray}
where $\varphi^{+}(\vec{x})$ and $\varphi^{-}(\vec{x})$ are two component spinors written in terms of the spherical harmonics $Y_{\ell m}(\theta, \vartheta)$ and defined as [11]
\begin{eqnarray}
\varphi^{+}(\vec{x})=\begin{pmatrix}
\sqrt{\frac{\ell+m+1}{2\ell+1}}\,Y_{\ell m}(\theta, \vartheta)\\
\sqrt{\frac{\ell-m+1}{2\ell+1}}\,Y_{\ell m}(\theta, \vartheta)
\end{pmatrix}\,,\nonumber
\end{eqnarray}
and
\begin{eqnarray}
\varphi^{-}(\vec{x})=\begin{pmatrix}
\sqrt{\frac{\ell-m}{2\ell+1}}\,Y_{\ell m}(\theta, \vartheta)\\
-\sqrt{\frac{\ell+m}{2\ell+1}}\,Y_{\ell m}(\theta, \vartheta)
\end{pmatrix}\,,\nonumber
\end{eqnarray}
where $(\theta, \vartheta)$ are two of spherical coordinates $(r, \theta, \vartheta)$. Because of vanishing magnetic field, one can choose a complete set of observables as $(H, \vec{J}^{2}, J_{z}, \vec{S}^{2}, K)$ for a spherical symmetric electric field $\vec{E}=E(r)\hat{r}$ where $\hat{r}$ is unit vector. The operator $H$ in this set is the Hamiltonian given in Eq. (4), $\vec{J}$ is the total angular momentum $\vec{J}=\vec{L}+\vec{S}$, where $\vec{L}$ is the orbital angular momentum, and $\vec{S}=\frac{1}{2}\,\vec{\Sigma}$ is the spin operator. The operator $K$ is given as $K=\beta(1+\vec{\Sigma}.\vec{L})$ which satisfies the commutation relations $[H, K]=[\vec{J}, K]=0$.

One can write the first part of solution to Eq. (5) as $\psi^{+}=(\phi^{+}, \chi^{+})^{t}$, where
\begin{eqnarray}
&&\phi^{+}(\vec{x})=F^{+}(r)\varphi^{\eta}(\vec{x})\,,\nonumber\\
&&\chi^{+}(\vec{x})=iG^{+}(r)(\vec{\sigma}.\hat{r})\varphi^{\eta}(\vec{x})\,,
\end{eqnarray}
where the superscript $\eta$ indicates $(+)$ in this case. By using the following equalities [18]
\begin{subequations}
\begin{align}
\vec{\sigma}.\hat{r}\varphi^{\pm}(\vec{x})&=\varphi^{\mp}(\vec{x})\,,\\
\vec{\sigma}.\vec{L}\varphi^{\pm}(\vec{x})&=\left(\vec{J}^{2}-\vec{L}^{2}-\frac{3}{4}\right)\varphi^{\pm}(\vec{x})\,,\nonumber\\
&=\begin{Bmatrix}
\ell  \\-\ell-1
\end{Bmatrix}\varphi^{\pm}(\vec{x})\,,\nonumber\\
&=\begin{Bmatrix}
-1+(j+1/2) \\ -1-(j+1/2)
\end{Bmatrix}\varphi^{\pm}(\vec{x})\,\,\,\ \mbox{for} j=\ell \pm \frac{1}{2}\,,\\
\vec{\sigma}.\vec{p}&=-i(\vec{\sigma}.\hat{r})\partial_{r}+\frac{i}{r}\,(\vec{\sigma}.\hat{r})(\vec{\sigma}.\vec{L})\,,
\end{align}
\end{subequations}
we obtain two first order differential equations from Eq. (5) with the help of Eq. (7) for the radial wave functions
\begin{subequations}
\begin{align}
\left(\frac{d}{dr}+\mu E(r)-\frac{\ell}{r}\right)F^{+}(r)&=-(\varepsilon+M)G^{+}(r)\,,\\
\left(\frac{d}{dr}-\mu E(r)+\frac{\ell+2}{r}\right)G^{+}(r)&=(\varepsilon-M)F^{+}(r)\,,
\end{align}
\end{subequations}
which gives the following because of the coordinate-dependence of mass
\begin{eqnarray}
\frac{1}{\varepsilon+M}\,\left\{-\frac{d^2}{dr^2}-\mu\,\frac{E(r)}{dr}+\frac{\ell}{r^2}+\frac{\frac{dM}{dr}}{\varepsilon+M}\,\left(\frac{d}{dr}+\mu E(r)-\frac{\ell}{r}\right)\nonumber\right.\\+\left.\frac{1}{r}\left[\mu E(r)r-(\ell+2)\right]\,\left(\frac{d}{dr}+\mu E(r)-\frac{\ell}{r}\right)\right\}F^{+}(r)=(\varepsilon-M)F^{+}(r)\,.
\end{eqnarray}

In addition to position dependency of the mass function, here, we want to write the mass function is also energy-dependent. For this aim, we tend to choose the option written as $\frac{dM}{dr}=(\varepsilon+M)^{2}$ which gives $M(r)=-1/r-\varepsilon+const.$. This gives also opportunities such as  writing Eq. (10) in a more simpler form and getting an analytical solution. By taking the 'integration constant' is zero, we have a second order differential equation for the radial wave function for $F^{+}(r)$
\begin{eqnarray}
&&\frac{d^{2}F^{+}(r)}{dr^2}+\frac{3}{r}\,\frac{dF^{+}(r)}{dr}\nonumber\\&+&\left[\mu\,\frac{dE(r)}{dr}-\mu^2E^2(r)+(3+2\ell)\mu\,\frac{E(r)}{r}-\frac{\ell(\ell+1)}{r^2}\right]F^{+}(r)=-[\varepsilon^{2}-M^{2}(r)]F^{+}(r)\,,\nonumber\\
\end{eqnarray}

The second part of solution to Eq. (5) as $\psi^{-}=(\phi^{-}, \chi^{-})^{t}$, where
\begin{eqnarray}
&&\phi^{-}(\vec{x})=F^{-}(r)\varphi^{\eta}(\vec{x})\,,\nonumber\\
&&\chi^{-}(\vec{x})=iG^{-}(r)(\vec{\sigma}.\hat{r})\varphi^{\eta}(\vec{x})\,,
\end{eqnarray}
where the superscript $\eta$ indicates $(-)$ in this case. With the help of Eq. (8), we write two first order differential equations
\begin{subequations}
\begin{align}
\left(\frac{d}{dr}+\mu E(r)-\frac{\ell+2}{r}\right)F^{-}(r)&=-(\varepsilon+M)G^{-}(r)\,,\\
\left(\frac{d}{dr}-\mu E(r)-\frac{\ell}{r}\right)G^{-}(r)&=(\varepsilon-M)F^{-}(r)\,,
\end{align}
\end{subequations}

We obtain the following for the position-dependent mass
\begin{eqnarray}
\frac{1}{\varepsilon-M}\left\{\frac{d^2}{dr^2}-\mu\frac{d E(r)}{dr}+\frac{\ell}{r^2}+\frac{\frac{dM}{dr}}{\varepsilon-M}\,\left(\frac{d}{dr}-\mu E(r)-\frac{\ell}{r}\right)+\,\nonumber\right.\\+\left.\frac{1}{r}\left[\mu E(r)r+\ell+2\right]\,\left(\frac{d}{dr}-\mu E(r)-\frac{\ell}{r}\right)\right\}G^{-}(r)=-(\varepsilon+M)G^{-}(r)\,.
\end{eqnarray}

By taking a similar option $\frac{dM}{dr}=(\varepsilon-M)^{2}$ ($M(r)=-1/r+\varepsilon+const.$), and setting the 'integration constant' to zero, we have for the radial function $G^{-}(r)$
\begin{eqnarray}
&&\frac{d^{2}G^{-}(r)}{dr^2}+\frac{3}{r}\,\frac{dG^{-}(r)}{dr}\nonumber\\&-&\left[\mu\,\frac{dE(r)}{dr}+\mu^2E^2(r)+(3+2\ell)\mu\,\frac{E(r)}{r}+\frac{\ell(\ell+1)}{r^2}\right]G^{-}(r)=-[\varepsilon^{2}-M^{2}(r)]G^{-}(r)\,,\nonumber\\
\end{eqnarray}

In the following, we try to solve Eq. (11) and Eq. (15) analytically for a given central electric field configuration $E(r)$ within the position-dependent mass formalism, i.e., $M \rightarrow M(r)$. But first, we shall give a brief summary of the NU method in the next section.

\section{T\lowercase{he} N\lowercase{ikiforov}-U\lowercase{varov} M\lowercase{ethod}}

Let us consider a second order differential equation with the following form [19]
\begin{eqnarray}
\left[\nu^2(r)\,\frac{d^2}{dr^2}+\nu(r)\tilde{\tau}(r)\,\frac{d}{dr}+\tilde{\nu}(r)\right]y(r)=0\,,
\end{eqnarray}
Here $\nu(r)$, and $\tilde{\nu}(r)$ are polynomials of $r$, at most, second degree, and $\tilde{\tau}(r)$ is a first degree polynomial. We take the ansatz for a particular solution
\begin{eqnarray}
y(r)=h(r)g(r)\,,
\end{eqnarray}
Inserting into Eq. (16) gives
\begin{eqnarray}
\left[\nu(r)\,\frac{d^2}{dr^2}+\tau(r)\,\frac{d}{dr}+\lambda\right]g(r)=0\,,
\end{eqnarray}
where $\lambda$ will be determined. The solution of Eq. (16) can be written by using Rodriguez formula
\begin{eqnarray}
g_{n}(r) \sim \frac{1}{\rho(r)}\,\frac{d^{n}}{dr^{n}}\left[\nu^{n}(r)\rho(r)\right]\,,
\end{eqnarray}
with the weight function $\rho(r)$ satisfying [14]
\begin{eqnarray}
\rho(r)\,\frac{d\nu(r)}{dr}+\nu(r)\,\frac{d\rho(r)}{dr}=\rho(r)\tau(r)\,,
\end{eqnarray}

The other part of whole solution in Eq. (17) is defined in terms of a required function $\pi(r)$ as
\begin{eqnarray}
\frac{1}{h(r)}\,\frac{dh(r)}{dr}=\frac{\pi(r)}{\nu(r)}\,,
\end{eqnarray}
The polynomial $\pi(r)$ is given within the method as
\begin{eqnarray}
\pi(r)=\frac{1}{2}\left[\nu\,'(r)-\tilde{\tau}(r)\right]\mp \left[\frac{1}{4}\left[\nu\,'(r)-\tilde{\tau}(r)\right]^2-
\tilde{\nu}(r)+k\nu(r)\right]^{1/2}\,,
\end{eqnarray}
where prime denotes the derivative with respect to $r$. The parameter introduced in Eq. (18) and $k$ in the above equation satisfy
\begin{eqnarray}
\lambda=k+\pi\,'(r)\,,
\end{eqnarray}
Since square root in the polynomial $\pi(r)$ in Eq. (22) must be a square then this defines the constant $k$. Replacing $k$ into Eq. (22), we define
\begin{eqnarray}
\tau(r)=\tilde{\tau}(r)+2\pi(r)\,,
\end{eqnarray}
The polynomial $\tau(r)$ must have a negative derivative in order to satisfy $\rho(r)>0$, and $\nu(r)>0$. This leads to the choice of the solution. If $\lambda$ satisfies the condition
\begin{eqnarray}
\lambda=\lambda_{n}=-n\tau\,'(r)-\frac{1}{2}\,n(n-1)\nu\,''(r)\,,\,\,\,n=0, 1, 2, \ldots
\end{eqnarray}
then the hypergeometric type equation (16) has a particular solution with degree $n$.

\section{T\lowercase{he} B\lowercase{ound} S\lowercase{tates}}

We find the energy eigenvalues of the Dirac-Pauli equation for a neutral spin-$\frac{1}{2}$ particle moving in a electromagnetic field having the form $E(r)=a+b/r$. Such a configuration for electric field can be produced by an infinite line of charge with a constant charge per unit length $\zeta$ [20]. Inserting it into Eq. (11), taking the mass function given above, we have
\begin{eqnarray}
\frac{d^{2}F^{+}(r)}{dr^2}+\frac{3}{r}\,\frac{dF^{+}(r)}{dr}+\frac{1}{r^2}\left[-A^2_{1}r-A^2_{2}r^{2}-A^2_{3}\right]F^{+}(r)=0\,,
\end{eqnarray}
where
\begin{eqnarray}
&&-A^2_{1}=\mu a(3+2\ell)-2\mu^{2}ab-2\varepsilon\,,\nonumber\\
&&-A^2_{2}=-\mu^2 a^{2}\,,\nonumber\\
&&-A^2_{3}=(1+\ell)(2\mu b-\ell-1)-\mu^{2}b^{2}\,,
\end{eqnarray}
We compare Eq. (26) with Eq. (16) for using the NU method, then we have
\begin{eqnarray}
\tilde{\tau}(z)=3\,\,;\nu(r)=r\,\,;\tilde{\nu}(z)=-A^2_{1}r-A^2_{2}r^{2}-A^2_{3}\,,
\end{eqnarray}
We obtain the following for the polynomial $\pi(r)$ from Eq. (22)
\begin{eqnarray}
\pi(z)=-1\pm \sqrt{A^2_{2}r^2+(k+A^2_{1})r+1+A^2_{3}\,}\,,
\end{eqnarray}
We define the constant $k$ by setting the discriminant of the expression under the square root to zero which gives
$k=-A^2_{1}-2A_{2}\,\sqrt{1+A^2_{3}\,}$. We replace it into above equation, and obtain
\begin{eqnarray}
\pi(r)=-1+\sqrt{1+A^2_{3}\,}-A_{2}r\,,\nonumber
\end{eqnarray}
which gives from Eq. (24)
\begin{eqnarray}
\tau(r)=1+2\left(\sqrt{1+A^2_{3}\,}-A_{2}r\right)\,,
\end{eqnarray}
with a negative derivative. The constant $\lambda$ in Eq. (23) becomes $\lambda=-A^2_{1}-A_{2}[1+2L(\ell)]$ with $L(\ell)=\sqrt{1+A^2_{3}\,}$, and $\lambda_{n}$ in Eq. (25) becomes $\lambda_{n}=2nA_{2}$. Substituting the values of parameters given in Eq. (27), and setting $\lambda=\lambda_{n}$, we find the energy eigenvalues
\begin{eqnarray}
&&\varepsilon=-\mu a[n-\ell-1+L(\ell)+\mu b]\,.
\end{eqnarray}
We observe that we have only particle solutions unless $n+L(\ell)+\mu b > 1+\ell$, while only antiparticle solutions for the other case. The eigenvalues depend on the quantum numbers $(n, \ell)$ which means that the degeneracy is finite for the Dirac-Pauli equation for the case where the mass is a function of spatially coordinate for the radially linear electric field. We summarize our numerical results, and dependency of energies on parameters $a$, and $b$ in Figs. 1 and 2, respectively. We prefer to give the variation on these parameters because the electric field is the source of the present problem. The Figs. 1 and 2 show that the energy eigenvalues linearly increase while the parameter $a$ increases, and decrease linearly while the value of parameter $b$ increases. The increasing with $a$, and decreasing with $b$ are expected because the parameter $a$ represents the linear dependency, and the other one represents the inverse-linear dependency in electromagnetic configuration.

Now let us find the normalized wave functions. We first compute the weight function from Eq. (20) and Eq. (30) as
\begin{eqnarray}
\rho(r) \sim r^{2L(\ell)}e^{-2A_{2}r}\,,\nonumber
\end{eqnarray}
and Eq. (19) gives
\begin{eqnarray}
g_{n}(r) \sim r^{-2L(\ell)}e^{2A_{2}r}\frac{d^{n}}{dr^{n}}\left[z^{n+2L(\ell)}e^{-2A_{2}r}\right]\,,\nonumber
\end{eqnarray}
which can be written in terms of the generalized Laguerre polynomials [21]
\begin{eqnarray}
g_{n}(r) \sim L_{n}^{2L(\ell)}(2A_{2}r)\,.
\end{eqnarray}
We write the other part of solution from Eq. (21)
\begin{eqnarray}
h(r) \sim r^{L(\ell)-1}e^{-A_{2}r}\,,
\end{eqnarray}
Thus, the whole wave functions $F^{+}(r)$ become
\begin{eqnarray}
F^{+}(r)=Nr^{L(\ell)-1}e^{-A_{2}r}L_{n}^{2L(\ell)}(2A_{2}r)\,,
\end{eqnarray}
and Eq. (9) gives the other component as
\begin{eqnarray}
G^{+}(r)=Nr^{L(\ell)-1}e^{-A_{2}r}\left\{(L(\ell)-1-A_{2}r)L_{n}^{2L(\ell)}(2A_{2}r)+r\frac{d}{dr}L_{n}^{2L(\ell)}(2A_{2}r)\right\}\,.
\end{eqnarray}
By using the normalization condition
\begin{eqnarray}
\int_{0}^{\infty}\left\{[F^{\pm}(r)]^2+[G^{\pm}(r)]^2\right\}r^2dr=1\,,\nonumber
\end{eqnarray}
and after cumbrous calculations we give the normalization constant
\begin{eqnarray}
N=\left[\frac{(2A_{2})^{1+2L(\ell)}}{\frac{[n+2L(\ell)]!(1+\omega^{2}_{1})}{n!}+\frac{[n-1+2L(\ell)]!\omega^{2}_{2}}{(n-1)!}+\frac{[n+1+2L(\ell)]!\omega^{2}_{3}}{(n+1)!}}\right]^{1/2}\,,
\end{eqnarray}
where
\begin{eqnarray}
\omega_{1}=\omega'_{1}+\omega'_{2}[2n+2L(\ell)+1]; \omega_{2}=[n+2L(\ell)](1+\omega'_{2});\omega_{3}=\omega'_{2}(1+n)\,,\nonumber
\end{eqnarray}
with
\begin{eqnarray}
\omega'_{1}=n-1-\ell+L(\ell)+\mu b;\omega'_{2}=\frac{\mu a}{2A_{2}}-\frac{1}{2}\,.\nonumber
\end{eqnarray}
Here and from now on we use the following identities for the associated Laguerre polynomials [21, 22]
\begin{eqnarray}
\frac{d}{dz}L_{n}^{k}(z)=\frac{1}{z}\left[nL_{n}^{k}(z)-(n+k)L_{n-1}^{k}(z)\right]\,,\nonumber\\
(n+k)L_{n-1}^{k}(z)+(n+1)L_{n+1}^{k}(z)=(2n+k+1-z)L_{n}^{k}(z)\,.\nonumber
\end{eqnarray}

We study the second part of solutions given in Eq. (15). Inserting the electric field expression, and taking the mass distribution obtained above, we have
\begin{eqnarray}
\frac{d^{2}G^{-}(z)}{dz^2}+\frac{3/2}{z}\,\frac{dG^{-}(z)}{dz}+\frac{1}{z^2}\left[-B^2_{1}z-B^2_{2}z^{2}-B^2_{3}\right]G^{-}(z)=0\,,
\end{eqnarray}
where
\begin{eqnarray}
&&B^2_{1}=\mu a(3+2\ell)+2\mu^{2}ab-2\varepsilon\,,\nonumber\\
&&B^2_{2}=\mu^2 a^{2}\,,\nonumber\\
&&B^2_{3}=(1+\ell)(2\mu b-\ell-1)+\mu^{2}b^{2}\,,
\end{eqnarray}
Following the same steps, we write the energy levels
\begin{eqnarray}
&&\varepsilon=\mu a[n+\ell+2+L(\ell)+\mu b]\,.
\end{eqnarray}
We observe that the solutions give only the particle states, and the energy levels depend also on the quantum number $\ell$ which means that the degeneracy is finite as before. The corresponding wave functions are
\begin{eqnarray}
G^{-}(r)=Nr^{L(\ell)-1}e^{-A_{2}r}L_{n}^{2L(\ell)}(2B_{2}r)\,,
\end{eqnarray}
and the other component is
\begin{eqnarray}
F^{-}(r)=Nr^{L(\ell)-1}e^{-B_{2}r}\left\{(L(\ell)-1-B_{2}r)L_{n}^{2L(\ell)}(2B_{2}r)+r\frac{d}{dr}L_{n}^{2L(\ell)}(2B_{2}r)\right\}\,.
\end{eqnarray}
The normalization constant is given for this case as
\begin{eqnarray}
N=\left[\frac{(2B_{2})^{1+2L(\ell)}}{\frac{[n+2L(\ell)]!(1+\omega^{2}_{1})}{n!}+\frac{[n-1+2L(\ell)]!\omega^{2}_{2}}{(n-1)!}+\frac{[n+1+2L(\ell)]!\omega^{2}_{3}}{(n+1)!}}\right]^{1/2}\,,
\end{eqnarray}
where
\begin{eqnarray}
\omega_{1}=\omega'_{1}+\omega'_{2}[2n+2L(\ell)+1]; \omega_{2}=[n+2L(\ell)](1+\omega'_{2});\omega_{3}=\omega'_{2}(1+n)\,,\nonumber
\end{eqnarray}
with
\begin{eqnarray}
\omega'_{1}=n-(1+\ell)+L(\ell)+\mu b;\omega'_{2}=\frac{1}{2}(\frac{\mu a}{B_{2}}-1)\,.\nonumber
\end{eqnarray}

One can see that it is also possible to study the bound state solutions of the Dirac-Pauli equation for a neutral spin-$\frac{1}{2}$ particle in a central electric field by using the NU method within the position dependent mass formalism. The scattering states for the same configuration of the electric field within the same formalism will be not studied here.

\section{C\lowercase{onclusion}}

We have studied the Dirac-Pauli equation of a neutral spin-$\frac{1}{2}$ particle coupled to a central electric field by non-minimal coupling by using the Nikiforov-Uvarov method within the position-dependent mass formalism. Solving the above equation and finding exact solutions by a different method could give another viewpoints within relativistic quantum mechanics. We have found out two second order differential equations, which have been obtained from four coupled first order equations, for both two different kinds of solutions. All bound state solutions are given in closed forms for a specific electric field configuration for which both negative- and positive-energy states are possible under specific constraints. We have obtained energy eigenvalue equations which involve only $\varepsilon$, not it's squared because of the setting the mass distribution as $dM(r)/dr=[\varepsilon \pm M(r)]^{2}$. As a result, we have reached both particle and antiparticle solutions for the upper component of the Dirac spinor together with some constraints on quantum numbers, while only particle solutions for lower component.

\section{A\lowercase{cknowledgments}}
One of authors (A.A.) thanks Prof Dr Andreas Fring from City University London and the Department of Mathematics for hospitality. This research was partially supported by the Scientific and Technical Research Council of Turkey and through a fund provided by University of Hacettepe. The authors would like to thank the referee for her/his kind suggestions.

\newpage

\begin{figure}
\centering \subfloat[][The ground-state energy.]{\includegraphics[height=2.2in,
width=3in, angle=0]{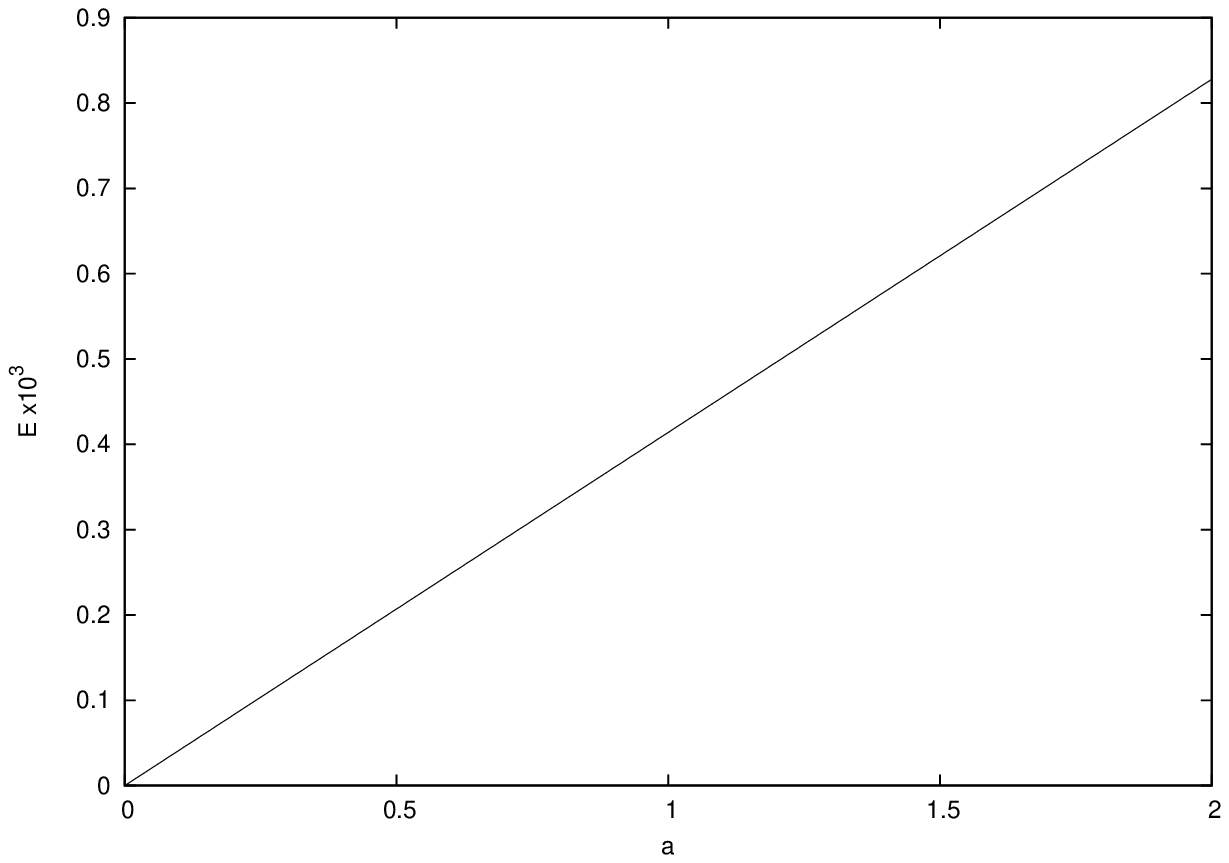}}
\subfloat[][First excited state energy with $\ell=0$.]{\includegraphics[height=2.2in, width=3in,
angle=0]{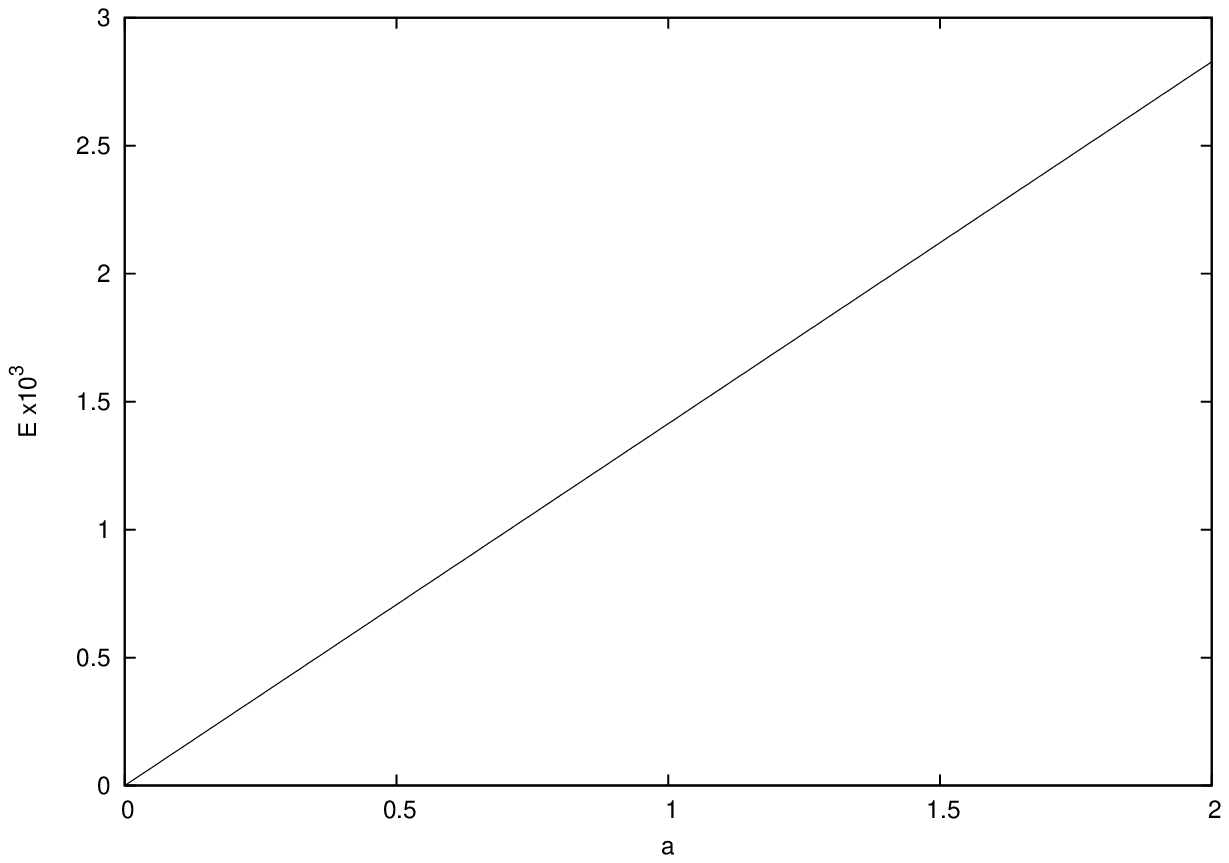}}\\
 \subfloat[][First excited state energy with $\ell=1$.]{\includegraphics[height=2.2in,
width=3in, angle=0]{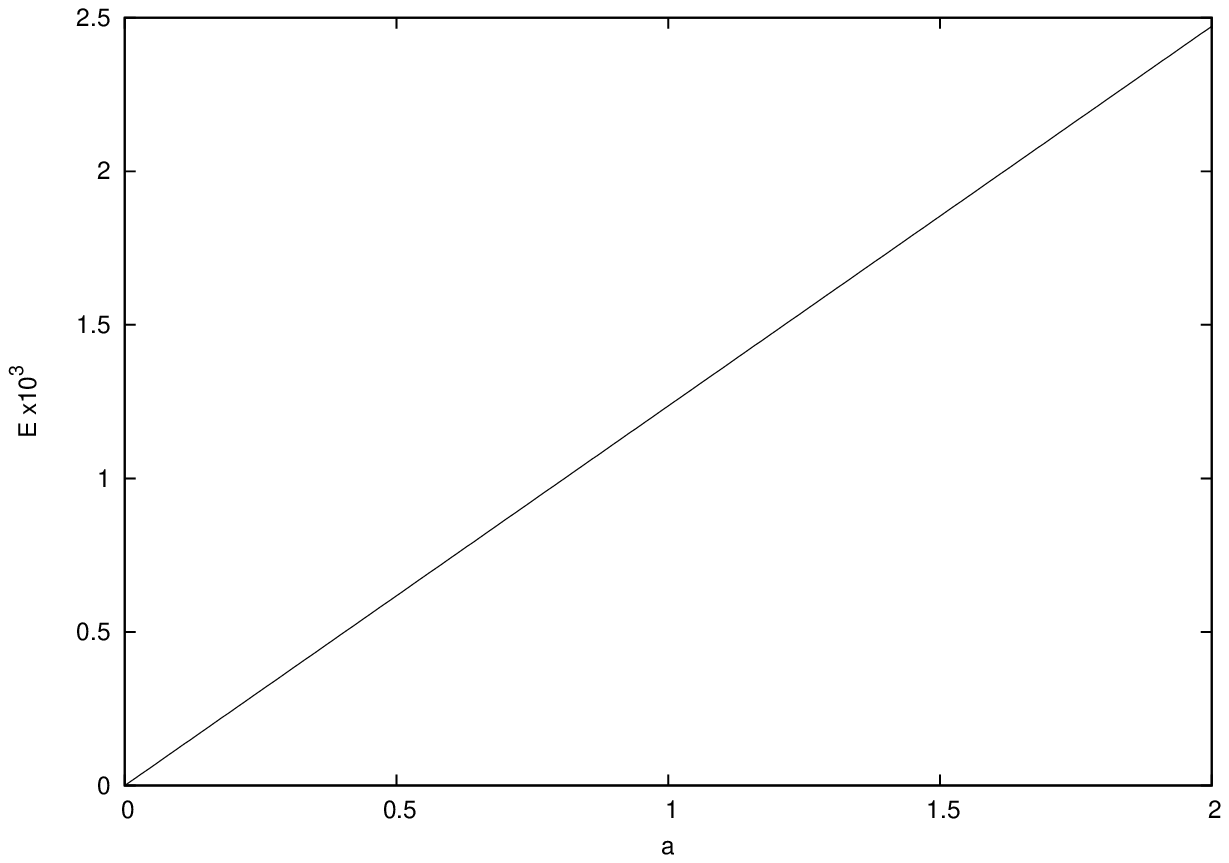}}
 \caption{The variation of energy with parameter $a$ for $b=1$, and $\mu=-0.001$.}
\end{figure}

\newpage

\begin{figure}
\centering \subfloat[][The ground-state energy.]{\includegraphics[height=2.2in,
width=3in, angle=0]{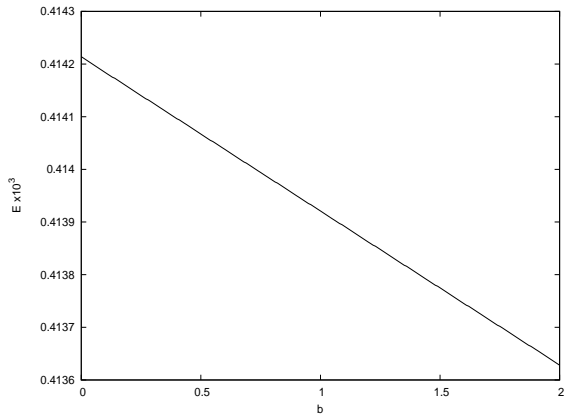}}
\subfloat[][First excited state energy with $\ell=0$.]{\includegraphics[height=2.2in, width=3in,
angle=0]{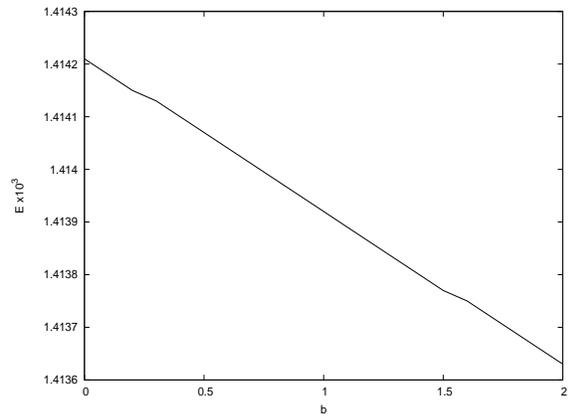}}\\
 \subfloat[][First excited state energy with $\ell=1$.]{\includegraphics[height=2.23in,
width=3in, angle=0]{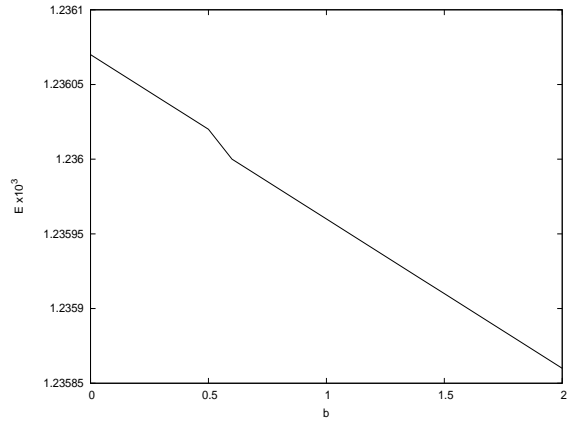}}
 \caption{The dependence of energy on the parameter $b$ for $a=1$, and $\mu=-0.001$.}
\end{figure}

\end{document}